\documentclass[10pt,conference]{IEEEtran}
\usepackage{graphicx}
\usepackage{tipa}
\usepackage{bbm}
\usepackage{mathrsfs}
\usepackage{amsfonts}
\usepackage{cite,url,subfigure,epsfig,graphicx}
\usepackage{amssymb,amsmath}
\usepackage{indentfirst}
\usepackage{algorithmic}
\usepackage{algorithm}
\usepackage{pdfpages}
\usepackage{multirow}
\usepackage{cases}
\usepackage{cite,url,subfigure,epsfig,graphicx}
\usepackage{times,verbatim,amsfonts,amsmath,color}

\begin{document}

\title{A Learning-based Honeypot Game for Collaborative Defense in UAV Networks}
\author{
\IEEEauthorblockN{Yuntao~Wang\IEEEauthorrefmark{2}, Zhou~Su\IEEEauthorrefmark{2}, Abderrahim~Benslimane\IEEEauthorrefmark{3}, Qichao~Xu\IEEEauthorrefmark{5}, Minghui~Dai\IEEEauthorrefmark{4}, and Ruidong~Li\IEEEauthorrefmark{9}}\\
\IEEEauthorblockA{
\IEEEauthorrefmark{2}School of Cyber Science and Engineering, Xi'an Jiaotong University, China\\
\IEEEauthorrefmark{3}Laboratory of Computer Sciences, Avignon University, France\\
\IEEEauthorrefmark{5}School of Mechatronic Engineering and Automation, Shanghai University, China\\
\IEEEauthorrefmark{4}State Key Laboratory of Internet of Things for Smart City, University of Macau, Macau, China\\
\IEEEauthorrefmark{9}Department of Electrical and Computer Engineering, Kanazawa University, Japan\\
}}

\maketitle

\begin{abstract}
The proliferation of unmanned aerial vehicles (UAVs) opens up new opportunities for on-demand service provisioning anywhere and anytime, but it also exposes UAVs to various cyber threats. Low/medium-interaction honeypot is regarded as a promising lightweight defense to actively protect mobile Internet of things, especially UAV networks.
Existing works primarily focused on honeypot design and attack pattern recognition, the incentive issue for motivating UAVs' participation (e.g., sharing trapped attack data in honeypots) to collaboratively resist distributed and sophisticated attacks is still under-explored.
This paper proposes a novel game-based collaborative defense approach to address optimal, fair, and feasible incentive mechanism design, in the presence of network dynamics and UAVs' multi-dimensional private information (e.g., valid defense data (VDD) volume, communication delay, and UAV cost).
Specifically, we first develop a honeypot game between UAVs under both partial and complete information asymmetry scenarios. We then devise a contract-theoretic method to solve the optimal VDD-reward contract design problem with partial information asymmetry, while ensuring truthfulness, fairness, and computational efficiency.
Furthermore, under complete information asymmetry, we devise a reinforcement learning based distributed method to dynamically design optimal contracts for distinct types of UAVs in the fast-changing network.
Experimental simulations show that the proposed scheme can motivate UAV's collaboration in VDD sharing and enhance defensive effectiveness, compared with existing solutions.
\end{abstract}


\IEEEpeerreviewmaketitle

\section{Introduction}
With the advancements in communication and embedded technologies, unmanned aerial vehicles (UAVs) have been widely adopted in various applications such as crowd surveillance, telecommunications, disaster search, and medical delivery \cite{9696188,8946587,9035635}. Due to their low cost, flexible mobility, and on-demand deployment capacity, UAVs can be promptly dispatched to hard-to-reach sites (e.g., disaster areas) to undertake time-critical missions and provide urgent communication services utilizing line-of-sight (LoS) links. As UAVs are computer-controlled agents with wireless/radio interfaces, the widespread use of UAVs in service offering exposes them to a plethora of sophisticated cyber attacks \cite{8946587} such as hijacking, eavesdropping, denial-of-service (DoS), and data theft. 

To combat rising cyber threats, low/medium-interaction honeypots, as a supplemental active defense technique, offer a cost-effective alternative to enhance UAV defense. Honeypots are physical or virtual systems which imitate real devices to lure and trap attackers, allowing defensive designers to continuously learn new attack patterns.
Compared with resource-hungry high-interaction honeypots, low/medium-interaction honeypots (which emulate network operations on the TCP/IP stack) can offer lightweight defenses \cite{4267549} that are especially suited for mobile and resource-constrained devices (e.g., battery-powered UAVs), which have attracted various research efforts. 
For example, Vasilomanolakis \emph{et al}. \cite{Vasi2014HosTaGe} design a generic low-interaction honeypot prototype named HosTaGe for mobile devices to detect malicious wireless networks as they connect.
Meanwhile, Daubert \emph{et al}. \cite{Daub2018HoneyDrone} implement a medium-interaction honeypot prototype named HoneyDrone on small-size UAVs by simulating UAV-specific protocols in the honeypot.

Despite the fundamental contributions to system and software design of existing literature \cite{Vasi2014HosTaGe,Daub2018HoneyDrone}, the collaborative defense strategy for UAVs based on honeypots is rarely studied.
Particularly, with the current trend of distributed, sophisticated, and complex covert cyber attacks (e.g., advanced persistent threat (APT) and distributed DoS (DDoS)), there is a necessity for large-scale collaborative defense among UAVs for global situational awareness by exchanging trapped attack information (e.g., attack interaction logs) in local honeypots.
Nonetheless, as participating in such collaboration mechanisms not only incurs considerable costs (e.g., honeypot execution and communication costs) but also results in potential privacy leakage (e.g., UAV's configurations and flying route), UAVs may be reluctant to share their captured attack data without satisfactory incentives. Besides, malicious UAVs may disseminate dishonest attack information to mislead others. Hence, it is pressing to design an effective incentive mechanism to motivate UAVs to cooperate in the joint defense.

However, the following key challenges need to be addressed in designing such an incentive mechanism compatible with UAVs. Firstly, UAVs typically have multi-dimensional private information in terms of valid defense data (VDD) volume, communication delay, VDD cost, and privacy cost. The presence of UAV's multi-dimensional information asymmetry poses significant challenges in \emph{optimally} and \emph{fairly} distributing rewards to compensate for UAV costs.
Secondly, as UAV networks and attack behaviors can be highly dynamic, the shared defense data from UAVs should be timely aggregated to produce real-time defense strategies. As such, it remains a challenge to \emph{feasibly} implement the incentive mechanism in practical UAV applications with time-varying environments and stringent latency requirements.

\begin{figure}[!t]\setlength{\abovecaptionskip}{-0.1cm}
\centering
  \includegraphics[width=7.8cm]{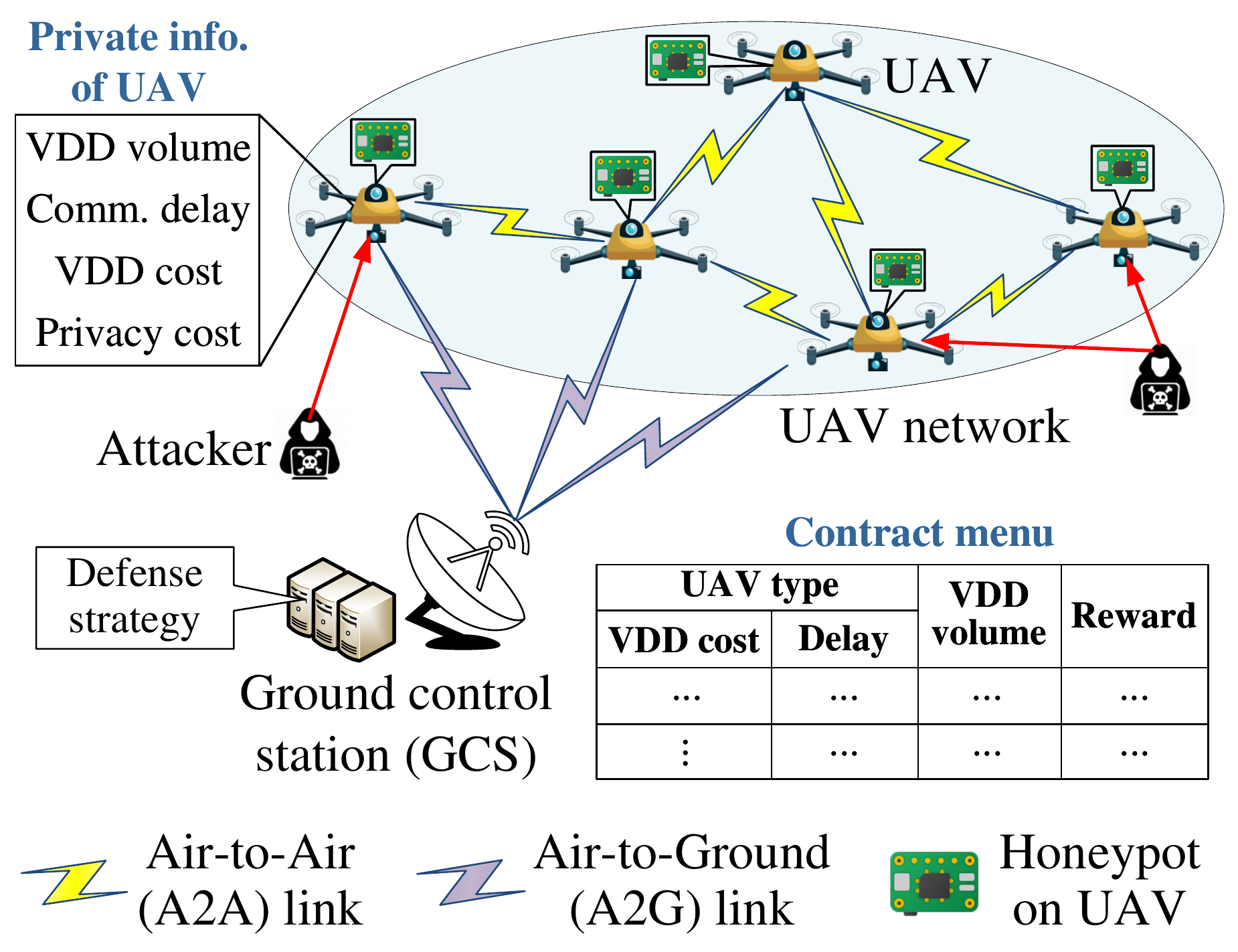}
  \caption{Illustration of the incentive mechanism for collaborative defense via sharing VDD in UAVs' local honeypots.}\label{fig:model}\vspace{-3.mm}
\end{figure}

This paper proposes a novel incentive-driven honeypot-based collaborative defense scheme for UAVs to improve defensive effectiveness, in which optimal, fair, and feasible incentives are offered to promote UAVs' honest cooperation in the face of information asymmetry and network dynamics.
Firstly, we present a UAV honeypot game framework consisting of multiple UAVs and a ground control station (GCS) serving as the network operator. In the game, the GCS designs a series of contracts (specifying the relation among VDD size, VDD cost, communication delay, and rewards) for heterogeneous UAVs, and each UAV chooses a contract to share its defense data.
Then, we formulate the optimal contract design problem for GCS under practical constraints and different levels of information asymmetry. Next, by leveraging the revelation principle, we analytically derive the optimal contract under \emph{partial information asymmetry} (i.e., GCS knows the numbers of different types of UAVs) and rigorously prove its truthfulness, fairness, and computational efficiency.
Furthermore, under \emph{complete information asymmetry} (i.e., GCS only knows the number of UAVs), we develop a two-tier reinforcement learning (RL) algorithm to dynamically acquire the optimal contracts for different types of UAVs via trials with high adaption to the fast-changing network environment.
Lastly, extensive simulations validate the efficiency of the proposed scheme in terms of UAV utility, defensive effectiveness, and convergence of dynamic contracts, in comparison with conventional approaches. 

The remainder of the paper is organized as follows. Section \ref{sec:RELATEDWORK} reviews the related works. Section \ref{sec:SYSTEMMODEL} formulates the system model and honeypot game model. Section \ref{sec:Solution} presents the optimal incentive mechanisms under partial and complete information asymmetry scenarios. Performance evaluation is given in Section \ref{sec:SIMULATION} and conclusions are drawn in Section \ref{sec:CONSLUSION}.

\section{Related Works}\label{sec:RELATEDWORK}

In the literature, various game-theoretical honeypot-based deception mechanisms have been proposed to enhance defense effectiveness.
Wang \emph{et al}. \cite{7857804} present a Bayesian game model between defenders and DDoS attackers in power grids, as well as a honeypot architecture to capture DDoS traffic on smart meters.
Tian \emph{et al}. \cite{9431233} present a game-based honeypot defense against APT attacks in industrial IoT, where stable strategies of attackers and defenders are derived under bounded rationality.
Wahab \emph{et al}. \cite{8675527} study a repeated Bayesian Stackelberg game model to detect smart attackers in the cloud, where attack patterns are learned from risky VMs using honeypots via support vector machine (SVM).
Tian \emph{et al}. \cite{9397770} investigate a contract game model to incentivize small-scale electricity suppliers (SESs) outfitted with honeypots to share defense data with power retailers for lower system defense costs.

However, the above works are built atop high-interaction honeypots on real hosts or VMs, which are inapplicable to UAV networks with high mobility and limited resources. In addition, both UAVs' multi-dimensional private information and different levels of information asymmetry are ignored in previous related works on incentive design.

\section{System Model and Game Formulation}\label{sec:SYSTEMMODEL}

\subsection{Network Model}\label{subsec:networkmodel}
Fig.~\ref{fig:model} depicts a typical honeypot-based collaborative defense scenario in a UAV network, consisting of one GCS (denoted as $G$) and $I$ flying UAVs. A group of UAVs (denoted as $\mathcal{I} = \{1,\cdots,I\}$) mounted with rich sensors are dispatched to a specific task area for immediate mission execution (e.g., power lines inspection). UAVs can exchange their flying information for collision avoidance via air-to-air (A2A) links. 
Besides, each UAV is equipped with a low/medium-interaction honeypot system to allow emulation, recording, and analysis of its captured malicious activities to mitigate cyber attacks. Let $S_i$ denote UAV $i$'s private valid defense data (VDD) volume, which means the data size of unknown attack interaction logs gathered by the UAV honeypot \cite{Daub2018HoneyDrone}. Here, UAVs are distinguished by their 2D private information: the marginal VDD cost and the communication delay. 
Let $\mathcal{J} = \{1,\cdots,J\}$ be the set of UAV types. We refer to a UAV with $\theta_j \triangleq (C_j,T_j)$ as a type-$j$ UAV. Here, $C_j$ means the unit cost for VDD generation, VDD transmission, and privacy loss of type-$j$ UAV. $T_j$ is the communication delay of type-$j$ UAV in transmitting VDD amount $S_j$ to the GCS. 

The GCS, as the coordinator of the UAV network, can communicate with UAVs via air-to-ground (A2G) links, perform UAV control (e.g., task assignment and trajectory planning), and carry out task data processing as well as security provisioning.
Traditionally, the GCS obtains defense data through external security service providers. In our scenario, the GCS additionally obtains defense data from UAVs, which have deployed the honeypot, for quicker attack recognition and better situational awareness.
To motivate UAVs' cooperation, the GCS offers a series of contracts $\Phi=\{T_{\max},\{\Phi_j\}_{j\in \mathcal{J}}\}$ including the maximum communication delay $T_{\max}$ (for all UAV types) and $J$ contract bundles $\{\Phi_j\}_{j\in \mathcal{J}}=\{S_j,R_j\}_{j\in \mathcal{J}}$ (one for each type). Here, $S_j$ and $R_j$ are the contributed VDD size and contractual reward (i.e., payment) of each type-$j$ UAV, respectively. For any UAV fails to deliver its VDD within $T_{\max}$, the GCS offers a zero-payment contract.

\emph{UAV Mobility Model.} Based on \cite{9696188}, the total time horizon is evenly divided into $T$ time slots with time length $\Delta_t$. UAV $i$'s instant 3D location at $t$-th time slot is denoted as ${\bf{l}}_i(t) = (x_i(t),y_i(t),z_i(t))$. The trajectories of UAVs are predetermined and controlled by the GCS $G$, which satisfies:
\begin{numcases}{}
{{\bf{l}}_i}(t+1) = {V_i}(t)\cdot{{\bf{w}}_i}(t) + {{\bf{l}}_i}(t),1 \le t \le T-1,\label{eq:mobility1} \hfill \\
{\rm{s.t.}}\; || {{\bf{l}}_i}(t+1) - {{\bf{l}}_i}(t) || \le \Delta_t V_{\max }^i. \label{eq:mobility2}
\end{numcases}
In (\ref{eq:mobility1}), ${V_i}(t)$ and ${\bf{w}}_i(t)$ are the flying velocity and trajectory direction of UAV $i$ at $t$-th time slot, respectively. In (\ref{eq:mobility2}), $V_{\max }^i$ is UAV $i$'s maximum velocity.


\emph{A2G Channel Model.} The A2G/G2A channel pathloss between UAV $i$ and GCS depends on the occurrence chances of LoS and non-LoS (NLoS) links \cite{8663615}, i.e.,
\begin{align}\label{eq:A2Gpathloss}
\Upsilon_{i,G}^{\mathrm{A2G}}(t)= 20 \log \big({\frac{4 \pi d_{i,G} \phi_{c}}{c}}\big) + {\Pr}_{\mathrm{{LoS}}} \kappa_{\mathrm{{L}}} + {\Pr}_{\mathrm{{NLoS}}} \kappa_{\mathrm{{N}}},
\end{align}
where $\kappa_{\mathrm{{L}}}$ and $\kappa_{\mathrm{{N}}}$ are additional attenuation factors for LoS and NLoS links, respectively.
$c$ is the speed of light, $\phi_{c}$ means the carrier frequency, and $d_{i,G}$ is the horizontal distance between UAV $i$ and the GCS. The LoS probability is a modified logistic function of the elevation angle $\theta_{i,G}=\arctan(\frac{z_i(t)-h_G}{d_{i,G}})$ \cite{8663615}, i.e.,
${\Pr}_{\mathrm{{LoS}}}= \left[ 1+ \iota_1\exp(-\iota_2(\theta_{i,G}-\iota_1))\right]^{-1}$.
Here, $h_G$ is the height of GCS, $\iota_1$ and $\iota_2$ are environment-related variables, and ${\Pr}_{\mathrm{{NLoS}}}=1-{\Pr}_{\mathrm{{LoS}}}$.

\subsection{Honeypot Game Model}\label{subsec:Honeypotmodel}
In the following, we define a UAV honeypot game as $\mathcal{G}=\{\{\mathcal{J},G\},\{T_{\max},\{S_j,R_j\}_{j\in \mathcal{J}}\},\{\{\mathcal{U}_j\}_{j\in \mathcal{J}},\mathcal{U}_G\}\}$.
\begin{itemize}
  \item \emph{Players.} The players in game $\mathcal{G}$ are (i) UAVs with diverse VDD-delay types in $\mathcal{J}$ and (ii) the GCS $G$.
  \item \emph{Strategies.} The strategy of the GCS is to determine the maximum communication latency $T_{\max}$ and design a set of feasible VDD-reward contracts $\{S_j,R_j\}_{j\in \mathcal{J}}$ to optimize its overall payoff. The strategy of each UAV is to select an optimal contract item for maximized payoff. 
  \item \emph{Payoffs.} The payoffs (or utilities) of each type-$j$ UAV and the GCS are denoted as $\mathcal{U}_j$ and $\mathcal{U}_G$, respectively.
\end{itemize}

\underline{Utility of UAV.} The utility of type-$j$ UAV that selects the contract item $\Phi_j=\{S_j,R_j\}$ is the revenue minuses its cost:
\begin{align}\label{eq:utility-UAV}
\mathcal{U}_{j}\left( \Phi_j \right) \!=\! \left\{ \begin{array}{ll}
	R_j\!-\!{C_j^1}S_j\!-\!{C_j^2}S_j\!-\!{C_0}, &if\, T_j\le T_{\max};\\
	-{C_j^1}S_j-{C_j^2}S_j-{C_0},& if\, T_j> T_{\max}.
\end{array} \right.
\end{align}

In (\ref{eq:utility-UAV}), ${C_j^1}$ is the unit cost of VDD creation and transmission, which is related to UAV's honeypot and communication capabilities. ${C_j^2}$ is UAV $i$'s unit privacy cost of VDD sharing. Both ${C_j^1}$ and ${C_j^2}$ are UAV's private information. Here, ${C_j}={C_j^1}+{C_j^2}$.
$C_0$ is the honeypot deployment cost of UAV. Besides, we have
\begin{align}\label{eq:latency}
T _{j}= \frac{S_j}{{B_j} \log_2\left(1 + \frac{1}{\sigma^2}P_j^{\mathrm{Tr}} 10^{-\frac{1}{10}\Upsilon_{j,G}^{\mathrm{A2G}}} \right)} ,
\end{align}
where ${B_j}$ is UAV $j$'s spectrum bandwidth, $\sigma^2$ means the power of Gaussian noise, and $P_j^{\mathrm{Tr}}$ is UAV $j$'s transmit power.
%

\underline{Utility of GCS.} The utility of the GCS is the overall satisfaction of cooperative defense minuses its total payments:
\begin{align}\label{eq:utility-GCS}
\mathcal{U}_{G}( \Phi )= \sum\limits_{j \in \mathcal{J}} \varpi \frac{N_j}{T_j}\log \left(1 + \mathbbm{1}_{T_j\le T_{\max}} S_j \right) - \mathbbm{1}_{T_j\le T_{\max}} N_j R_j.
\end{align}

In (\ref{eq:utility-GCS}), the first term indicates the satisfaction part related to UAV's VDD size and communication latency. 
Based on \cite{9035635,5738226}, we utilize the natural logarithmic function for satisfaction modelling. $\varpi$ is a positive satisfaction factor. ${N_j}$ is the number of type-$j$ UAVs, which satisfies $\sum_{j\in \mathcal{J}}{N_j}=I$.
$\mathbbm{1}_{T_j\le T_{\max}}$ is an indicator function, whose value is one if ${T_j\le T_{\max}}$ holds; otherwise its value is zero.

\underline{Optimization Problem.} The solution of the game $\mathcal{G}$ is to design the \emph{optimal contracts}, i.e., $\Phi^*=\{T_{\max},\{S_j^*,R_j^*\}_{j\in \mathcal{J}}\}$, for all types of UAVs while enforcing \emph{contractual feasibility}, where the optimization problem is formulated as:
\begin{align}\label{eq:optproblem1}
\mathbf{Problem}~1:\mathop {\max }\nolimits_{\Phi}\, \mathcal{U}_{G}\left( \Phi \right)~~~~~~~~~~~~
\end{align}
\begin{numcases}{{~~\rm{s.t.}}}	
0\le S_{j}\le S_{\max},  \label{eq:cons1} \hfill \\
\mathcal{U}_{j}( \Phi_j ) \geq 0,\forall j  \in \mathcal{J}, \label{eq:cons2} \hfill \\
\mathcal{U}_{j}( \Phi_j ) \geq  \mathcal{U}_{j}( \Phi_{j'} ), \forall j,j' \in \mathcal{J}, j\ne j'. \label{eq:cons3} \hfill
\end{numcases}

Where constraint (\ref{eq:cons1}) means that the VDD volume is constrained by the upper bound $S_{\max}$ and lower bound $0$. Constraints (\ref{eq:cons2}) and (\ref{eq:cons3}) enforce the contractual feasibility \cite{9397770}. Constraint (\ref{eq:cons2}) ensures the \emph{individual rationality (IR)}, i.e., each type-$j$ UAV receives a non-negative utility if it truthfully selects the contract item designed for its type. Constraint (\ref{eq:cons3}) ensures the \emph{individual compatibility (IC)}, i.e., each type-$j$ UAV prefers the contract item designed for its type instead of others.

\section{Solution of the UAV Honeypot game}\label{sec:Solution}
In this section, we design the optimal contracts for UAVs under both partial and complete information asymmetry scenarios. 
\begin{itemize}
  \item \emph{Partial information asymmetry scenario}. The GCS has prior knowledge of the total number of UAVs (i.e., $I$) and the specific number of each UAV type (i.e., $N_j,\forall j \in \mathcal{J}$), but is unaware of which UAV belongs to which type.
  \item \emph{Complete information asymmetry scenario}. The GCS only knows the number of UAVs and the number of UAV types. 
\end{itemize}
\subsection{Optimal Contract Design in Partial Information Asymmetry}\label{sec:StaticSolution}
Notably, there are $J^2$ IR and IC constraints in Problem 1, making it difficult to directly solve Problem 1, especially when $J$ is large. In what follows, IR and IC constraints are first transformed into a smaller number of equivalent ones using Lemma 1 and Theorem 1.
Let $\mathcal{J}'$ be the set of UAV types with $\mathbbm{1}_{T_j\le T_{\max}}=1$, i.e., $\mathcal{J}'=\{j|{T_j\le T_{\max}}\}$. Then we reindex UAV types in $\mathcal{J}'$ in descending order of the marginal VDD cost, i.e., $C_1>C_2>\cdots>C_{J'}$, where $J'=|\mathcal{J}'|$.

\emph{Lemma 1:} If IC constraints in (\ref{eq:cons3}) are met for all UAV types, then IR constraints in (\ref{eq:cons2}) can be replaced by $\mathcal{U}_{1}( \Phi_1 ) \geq 0$.

\begin{IEEEproof}
The detailed proof can refer to Appendix A of the technical report \cite{OnlineAppendix} due to page limits.
\end{IEEEproof}


\emph{Theorem 1:} A contract $\Phi=\{T_{\max},\{\Phi_j\}_{j\in \mathcal{J}}\}$ is feasible if and only if it meets the following conditions:\vspace{-2mm}
\begin{enumerate}
  \item $\forall j \notin \mathcal{J}'$, $S_{j}=R_{j}=0$.
  \item $\forall j \in \mathcal{J}'$, the following three conditions hold:
  \begin{numcases}{}	
     0\le S_1 \le \cdots \le S_{J'}\, \&\, 0\le R_1 \le \cdots \le R_{J'},~~~~\ \label{eq:theo1cons1} \hfill \\
     R_1-{C_1}S_1-{C_0}\ge 0,  \label{eq:theo1cons2} \hfill \\
     \begin{gathered}
     C_{j}\left( S_j - S_{j-1} \right) \le R_{j} - R_{j-1}~~~~~~~~~~~~~~~~~~~ \\~~~~~~~~\le C_{j-1}\left( S_{j}-S_{j-1} \right),\ j = 2,\cdots,{J'}. \label{eq:theo1cons3} \hfill
     \end{gathered}
 \end{numcases}
\end{enumerate}

\begin{IEEEproof}
See Appendix B of the technical report \cite{OnlineAppendix}.
\end{IEEEproof}

\emph{Remark.}
For any UAV type $j \notin \mathcal{J}'$, Theorem 1 shows that the required contractual VDD size and reward are zero.
Constraint (\ref{eq:theo1cons1}) means that the GCS should demand more VDD from UAVs with smaller marginal costs and offer more rewards to them.
Constraint (\ref{eq:theo1cons2}) indicates that if the UAV with the highest marginal cost satisfies the IR constraint, then IR constraints hold for all types of UAVs.
Constraint (\ref{eq:theo1cons3}) implies that if IC constraint holds between type-$j$ and type-$(j\!-\!1)$ UAVs, it also holds between type-$j$ UAV and any other type of UAV.

In the following Theorem 2, we derive the optimal reward strategy $\mathbf{R}^*(\mathbf{S})$ given any monotonic VDD size sequence $\mathbf{S}$.

\emph{Theorem 2.} Given any VDD volume sequence $\mathbf{S} = \{S_{j}\}_{j \in \mathcal{J}'}$ meeting $0\le S_{1} \le \cdots \le S_{J'} \le S_{\max}$, the unique optimal reward strategy $\mathbf{R}^*=\{R_{j}^*\}_{j \in \mathcal{J}'}$ is attained by:
\begin{enumerate}
  \item $\forall j \notin \mathcal{J}'$, $R_{j}^*(\mathbf{S})=0$.
  \item $\forall j \in \mathcal{J}'$,
  \begin{align}\label{eq:4-1-OptimalPrice}
    R_{j}^{*}\left( \mathbf{S} \right) =\left\{ \begin{array}{l}
    	{C_j}{S_{j}} \!+\! \sum\nolimits_{k=1}^{j-1}{\left( C_{k}\!-\!C_{k+1}\right){S_{k}}} \!+\! C_0,\\ ~~~~~~~~~~~~~~~~~~~ j=2,\cdots,J';\\
    	{C_j}{S_{j}}+ C_0,\ ~~~~j=1.\\
    \end{array} \right.
    \end{align}
\end{enumerate}
\vspace{-2mm}
\begin{IEEEproof}
See Appendix C of the technical report \cite{OnlineAppendix}.
\end{IEEEproof}

Theorem 2 shows that the optimal reward is positively correlated with UAV's shared VDD size, which ensures contractual fairness. 
According to Theorems 1--2, the original Problem 1 can be rewritten as the following simplified problem:\vspace{-1mm}
\begin{align}\label{eq:optproblem2}
\begin{gathered}
  ~~~\mathbf{Problem}~2:~\mathop {\max }\nolimits_{\Phi}\, \mathcal{U}_{G}\left( \Phi \right)  \hfill \\
  ~~\mathrm{s.t.}\left\{ \begin{gathered}
  \mathrm{C1:}\ 0\le S_1 \le \cdots \le S_{J'} \le S_{\max}, \hfill \\
  \mathrm{C2:}\ R_1-{C_1}S_1-{C_0}= 0, \hfill \\
  \mathrm{C3:}\ R_{j} \!-\! {C_j}S_{j} \!=\! R_{j-1} \!-\! {C_j}S_{j-1},\forall j\!=\!2,\!\cdots\!,J'. \hfill \\
\end{gathered}  \right. \hfill \\
\end{gathered}
\end{align}\vspace{1mm}
%

\emph{Lemma 2.} The optimal contractual VDD size strategy can be obtained by solving the relaxed problem of Problem~2 without the monotonicity constraint C1, i.e.,\vspace{-1mm}
\begin{align}\label{eq:optproblem3}
S_{j}^* = &\arg \mathop {{\max} }\limits_{0 \le {S_{j}} \le S_{\max}}\Big(\varpi N_j\frac{1}{T_j}\log (1 + S_j )- {C_j}S_{j}\times \nonumber \\& \sum\nolimits_{k=j}^{J'}{N_k} + C_{j+1}S_{j}\sum\nolimits_{k=j+1}^{J'}{N_k}\Big),\forall j \in \mathcal{J}'.
\end{align}

\begin{IEEEproof}
See Appendix D of the technical report \cite{OnlineAppendix}.
\end{IEEEproof}

Notably, (\ref{eq:optproblem3}) is a single variable optimization problem.
Let $\Im(S_j)\!=\!\varpi \frac{N_j}{T_j}\log (1 + S_j ) - {C_j}S_{j}\sum\nolimits_{k=j}^{J'}{N_k} + C_{j+1}S_{j}\sum\nolimits_{k=j+1}^{J'}{N_k}$. As $\frac{\mathrm{d}^2\Im(S_j)}{\mathrm{d} S_{j}^2}=-\frac{\varpi N_j}{{T_j}\left( 1+S_{j} \right) ^2}<0$, $\Im(S_j)$ is strictly concave with respect to $S_{j}$. Thereby, the optimal VDD size strategy $S_{j}^*$ can be obtained by
\begin{align}\label{eq:OptimalSize1}
S_{j}^*=\min\left\{S_{\max},\max\left\{\tilde{S}_{j}^{*},0\right\}\right\},
\end{align}
where the point $\tilde{S}_{j}^{*}$ satisfies $\frac{\mathrm{d}\Im(S_j)}{{\mathrm{d} S_{j}}}\!=\! 0$, i.e.,\vspace{-0.5mm}
\begin{align}\label{eq:OptimalSize2}
~~\tilde{S}_{j}^{*}\!=\!\left\{ \begin{array}{l}
 \resizebox{.096\hsize}{!}{$\frac{\varpi }{T_j C_j}$}-1,~~~~~~~~~~~~~~~~~~~~~~~~~~~~~~j=J';\\
 \resizebox{.5\hsize}{!}{$\frac{\varpi {N_j}} {T_j (C_j\sum\limits_{k=j}^{J'}{N_k}-C_{j+1}\sum\limits_{k=j+1}^{J'}{N_k})}$}\!-\!1,~ j< J'.\\ 
\end{array} \right.
\end{align}

If $\mathbf{S}^*=\{S_{j}^*\}_{j\in \mathcal{J}'}$ is an non-decreasing sequence (i.e., C1 holds), then $\mathbf{S}^*$ is exactly the solution of Problem~2. Nevertheless, the monotonicity constraint C1 may not hold under the general distribution of UAV types. Based on \cite{5738226}, a dynamic VDD size assignment method is designed to cope with this issue through bunching and ironing. Specifically, for any decreasing sub-sequence $\{S_{m}^*,S_{m+1}^*,\cdots,S_{n}^*\}\subseteq \mathbf{S}^*$, all its elements are dynamically adjusted by solving the following single variable optimization problem:
\begin{align}\label{eq:OptSizeAssign}
S_{l}^*=\arg\mathop{\max}\limits_{S_{l}} \sum\nolimits_{l = m}^n {\Im(S_{l})}, \forall l =m,m+1,\cdots,n.
\end{align}\vspace{-3mm}

\emph{Remark.}
The above process in (\ref{eq:OptSizeAssign}) is repeated until all the sub-sequences in $\mathbf{S}^*$ obtained from (\ref{eq:OptimalSize1})--(\ref{eq:OptimalSize2}) are non-decreasing. After that, the optimal contracts $\Phi^*=\{T_{\max},\{S_j^*,R_j^*\}_{j\in \mathcal{J}}\}$ can be obtained for all types of UAVs.

\subsection{RL-based Dynamic Contract Design in Complete Information Asymmetry}\label{sec:DynamicSolution}
Under complete information asymmetry, both the GCS and UAVs can apply policy hill-climbing (PHC) learning, a model-free RL technique, to make optimal reward and VDD size strategies in dynamic contract design through trials without explicitly knowing the type distribution and UAVs' private parameters. For both sides, the strategy-making process can be modeled as a finite Markov decision process (MDP).

\underline{PHC-based Reward Strategy of GCS.} At $t$-th time slot, the system state vector ${\bf{W}}^t = \left(W_1^t,\cdots,W_{J'}^t\right)$ observed by the GCS contains the previous VDD size of each type of UAV, i.e., ${\bf{W}}^t = {\bf{S}}^{t-1}$. For simplicity, the GCS uniformly quantizes the reward strategy into $A+1$ levels, i.e., ${R}_j \in \mathcal{A}=\{\frac{a}{A}\cdot R_{\max}\}_{0 \le a \le A}$, where $R_{\max}$ is the maximum reward that the GCS pays to a UAV.
Let $\mathcal{{Q}}\left({\bf{W}}^t, {\bf{R}}^t\right)$ represent the GCS's Q-function (i.e., expected long-term discounted utility) of each state-action pair, which is updated based on iterative Bellman equation:
\begin{align}\label{eq:6-1}
\mathcal{{Q}}\left({W}_j^t, R_j^t\right) \leftarrow &(1-\kappa_1)\mathcal{{Q}}\left({W}_j^t, R_j^t\right) + \kappa_1 \left\{ {\mathcal{U}_G}\left({W}_j^t, R_j^t\right) \right. \nonumber \\
&\left. {+ \varphi_1 \mathop {\max }\limits_{R_j} \mathcal{{Q}}\left(W_j^{t+1}, R_j^{t+1}\right)} \right\}, \forall j \in \mathcal{J}',
\end{align}
where $\kappa_1$ is the learning rate, and $\varphi_1$ is the discount factor.

To tradeoff the exploration and exploitation in PHC, the mixed-strategy table $\pi\left({\bf{W}}^t, {\bf{R}}^t\right)$ is updated by increasing the chance of behaving greedily by a small value $\rho_1$, and lowering other chances by $-\frac{\rho_1}{A+1}$, i.e.,
\begin{align}\label{eq:6-3}
\pi &\left(W_j^t, R_j^t\right) \gets \pi \left(W_j^t, R_j^t\right)\nonumber \\
&+\left\{ \begin{array}{cl}
	\rho _1,&if\ R_j^{t}=\arg\max_{R_j}\mathcal{Q}\left( W_j^{t},R_j \right);\\
	-\frac{\rho _1}{A+1},&otherwise.
\end{array} \right.
\end{align}
The GCS opts its reward strategy $R_j^t,j\in \mathcal{J}'$ based on the above mixed-strategy table, i.e.,
\begin{align}\label{eq:6-4}
\Pr\big( R_j^{t} = \hat{R}_j \big) = \pi \big( W_j^t,\hat{R}_j \big) ,\forall \hat{R}_j\in \mathcal{A}.
\end{align}

\underline{PHC-based VDD Size Strategy of UAV.}
The state vector $\tilde{W}_j^t$ observed by type-$j$ UAV at $t$-th time slot contains the previous GCS's reward, i.e., $\tilde{W}_j^t = {R}_{j}^{t-1}$.
Each UAV uniformly quantizes its VDD size strategy into $B+1$ levels, i.e., $S_j \in \mathcal{B}=\{\frac{b}{B}\cdot S_{\max} \}_{0 \le b \le B}$. Let $\tilde{\mathcal{{Q}}}\left(\tilde{W}_j^t, S_j^t\right)$ denote the Q-function of type-$j$ UAV. Similarly, the Q-function is updated by:
\begin{align}\label{eq:6-5}
\tilde{\mathcal{{Q}}}\left(\tilde{W}_j^t, S_j^t\right) &\leftarrow (1-\kappa_2)\tilde{\mathcal{{Q}}}\left(\tilde{W}_j^t, S_j^t\right) + \kappa_2 \left\{ {\mathcal{U}_j}\left(\tilde{W}_j^t, S_j^t\right) \right. \nonumber \\
&\left. {+ \varphi_1 \mathop {\max }\limits_{S_j} \tilde{\mathcal{{Q}}} \left( \tilde{W}_{j}^{t+1}, S_{j}^{t+1} \right) } \right\},\forall j \in \mathcal{J}',
\end{align}
where $\kappa_2$ is the learning rate, and $\varphi_2$ is the discount factor.

The mixed-strategy table $\tilde{\pi}\left(\tilde{W}_{j}^t, S_{j}^t\right)$ in PHC is updated by
\begin{align}\label{eq:6-7}
\tilde{\pi}&\left(\tilde{W}_{j}^t, S_{j}^t\right) \gets \tilde{\pi}\left(\tilde{W}_{j}^t, S_{j}^t\right)\nonumber \\
&+\left\{ \begin{array}{cl}
	\rho _2,&if\ S_j^{t}=\arg\max_{S_j}\tilde{\mathcal{{Q}}}\left(\tilde{W}_{j}^t, S_j\right);\\
	-\frac{\rho _2}{B+1},&otherwise.
\end{array} \right.
\end{align}
According to the mixed-strategy table, each type-$j$ UAV ($j\in \mathcal{J}'$) chooses its VDD size strategy $S_{j}^t$ with the following chance:
\begin{align}\label{eq:6-8}
\Pr\left( S_j^{t} = \hat{S}_{j} \right) = \pi \left(\tilde{W}_{j}^t,\hat{S}_{j}\right) ,\forall \hat{S}_{j}\in \mathcal{B}.
\end{align}

\emph{Remark.}
The above two-tier strategy-making process is repeated between the GCS and each UAV in $\mathcal{J}'$ until both of their strategies converge to stable values. Besides, to speed up the convergence rate, a hotbooting technique is employed for both sides by learning from similar scenarios in an offline manner for efficient initialization of the Q-table and mixed-strategy table. 


\section{PERFORMANCE EVALUATION}\label{sec:SIMULATION}

\vspace{-0.05cm}
\subsection{Simulation Setup}\label{subsec:evalution1}\vspace{-0.05cm}
We consider a simulation area of $200\times200\times20\, \mathrm{m}^3$ with one GCS and $I=10$ uniformly distributed Parrot AR Drone 2.0 UAVs. Each UAV is embedded with a honeypot system and communicates with the GCS and other UAVs via Wi-Fi communications. The honeypot reads UAV profiles from the configuration file to emulate UAV's radio interfaces and offers low to medium interactions with adversaries for Wi-Fi protocols. The Telnet attack \cite{Daub2018HoneyDrone} is considered in the simulation, and the honeypot's captured VDD (including IP address, port number, connection type, commands, and timestamps of attackers) is recorded into a local MongoDB database. The GCS requests VDD from UAVs every $6$ seconds, with a maximum communication delay of $2$ seconds.
For simplicity, UAVs' types are assumed to be uniformly distributed. The lower and upper bounds of UAV's marginal VDD cost are set as $0.01$ and $1$, respectively.
For the utility model, we set $\varpi=6$, $C_0 = 1$, $T_{\max} = 2$ seconds, $S_{\max} = 300$ bytes.
For the PHC learning, we set $\kappa_1=\kappa_2=0.7$, $\varphi_1=\varphi_2=0.8$, $\rho_1=\rho_2=0.01$.

The following three conventional contract approaches are used for performance comparison with the proposed scheme.
\begin{itemize}
  \item \emph{Complete information contract}. In this ideal scenario, the GCS knows the private type of each UAV, and only IR constraints should be met in optimal contract design. The optimal contract under no information asymmetry satisfies:\\
  ~$\mathrm{1)}$ $\forall j \notin \mathcal{J}'$, $S_{j}=R_{j}=0$.
  \begin{numcases}{\mathrm{2)}\ \forall j \!\in\! \mathcal{J}'\!,\!}	
S_{j}^*\!=\!\min\big\{S_{\max},\max\big\{\frac{\varpi }{T_j C_j}\!-\!1,0\big\}\big\}\!,\!\  \label{eq:CIoptS} \hfill \\
R_{j}^*= {C_j} S_{j}^* + C_0. \label{eq:CIoptR}
\end{numcases}
  \item \emph{Linear contract}. The reward offered by the GCS is in direct proportion to UAV's shared VDD size in this contract, i.e., $R_j=\mu_{G} S_j + C_0$, $\forall j \in \mathcal{J}$, where $\mu_G$ is the unit payment per VDD size. Here, we set $\mu_G=C_1=1$.
  \item \emph{Uniform contract}. In this contract, the GCS applies a single uniform contract item for all types of UAVs, i.e., $\Phi_j=\{S_1^*,R_1^*\}$, $\forall j \in \mathcal{J}$.
\end{itemize}

\vspace{-0.05cm}
\subsection{Numerical Results}\label{subsec:evalution2}
\begin{figure*}[!tbp]\setlength{\abovecaptionskip}{-0.05cm}\vspace{-0.2cm}
\begin{minipage}[t]{0.243\textwidth}
\centering
    \includegraphics[height=3.37cm,width=\linewidth]{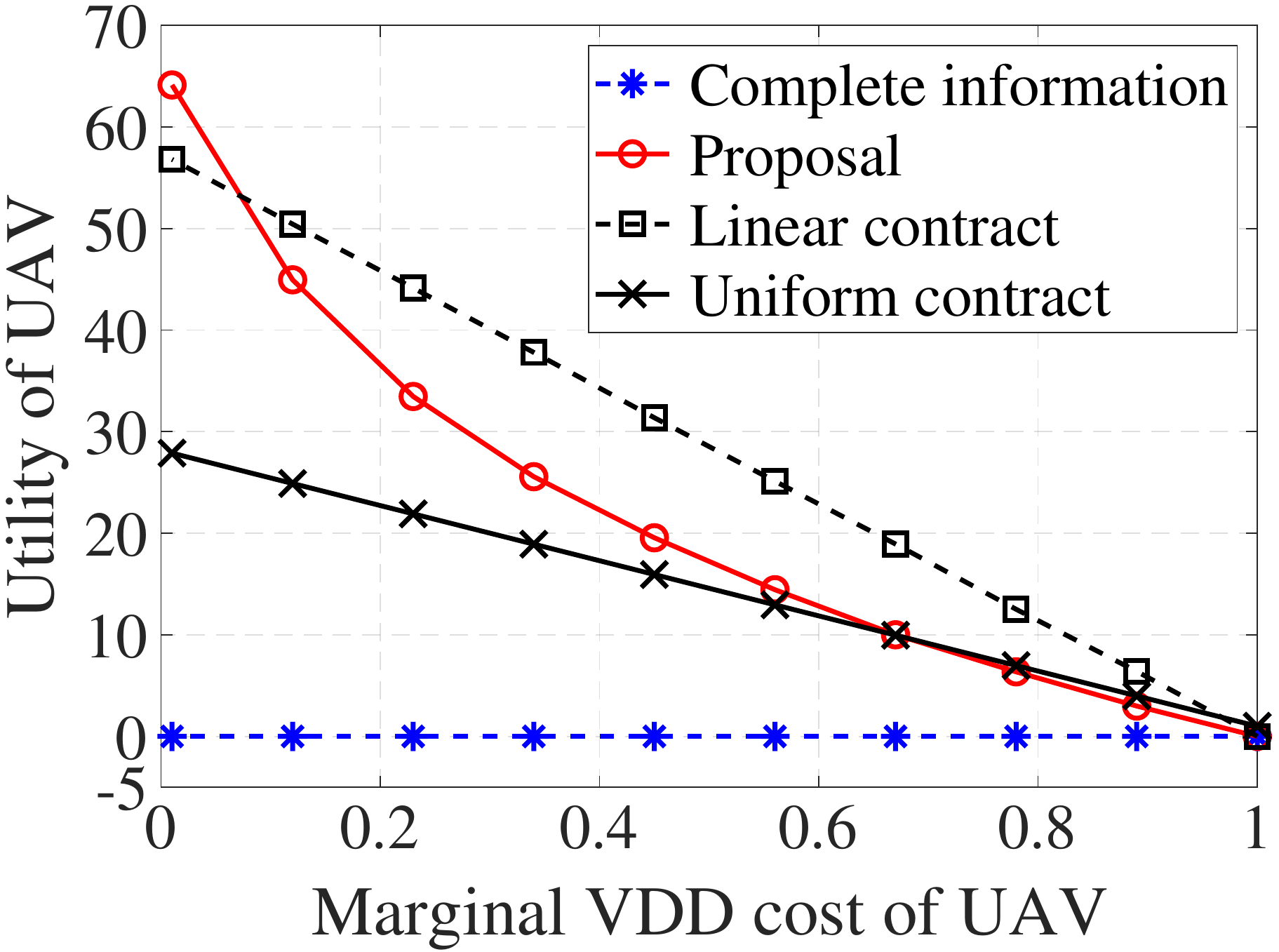}
    \caption{Utility of UAV vs. marginal VDD cost in the proposed scheme under partial information asymmetry, compared with other three contracts.}\label{fig:simu1}
\end{minipage}~~
\begin{minipage}[t]{0.243\textwidth}
\centering
    \includegraphics[height=3.37cm,width=\linewidth]{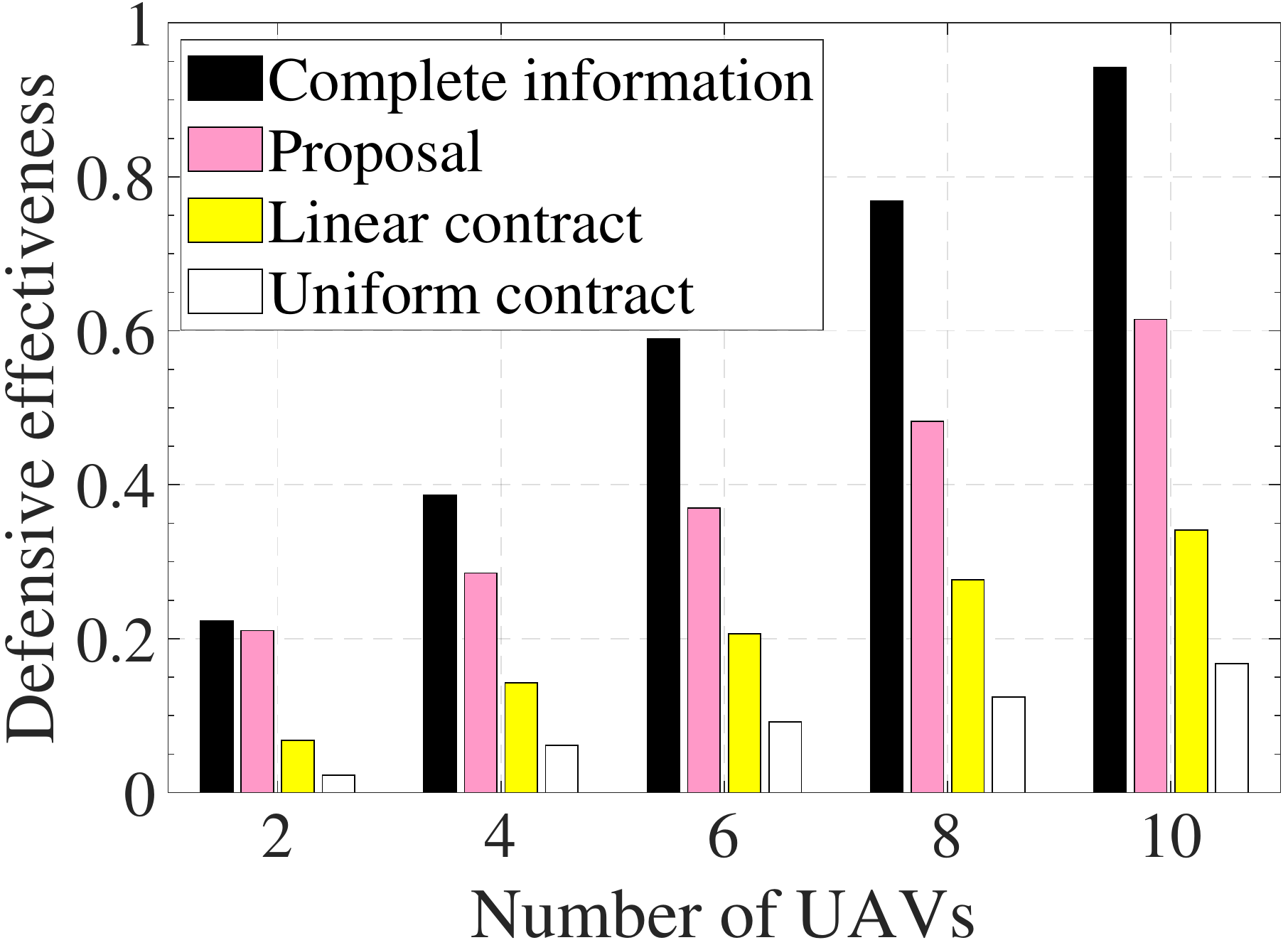}
    \caption{Defensive effectiveness vs. number of UAV in the proposed scheme under partial information asymmetry, compared with other three contracts.}\label{fig:simu2}
\end{minipage}~~
\begin{minipage}[t]{0.243\textwidth}
\centering
    \includegraphics[height=3.37cm,width=\linewidth]{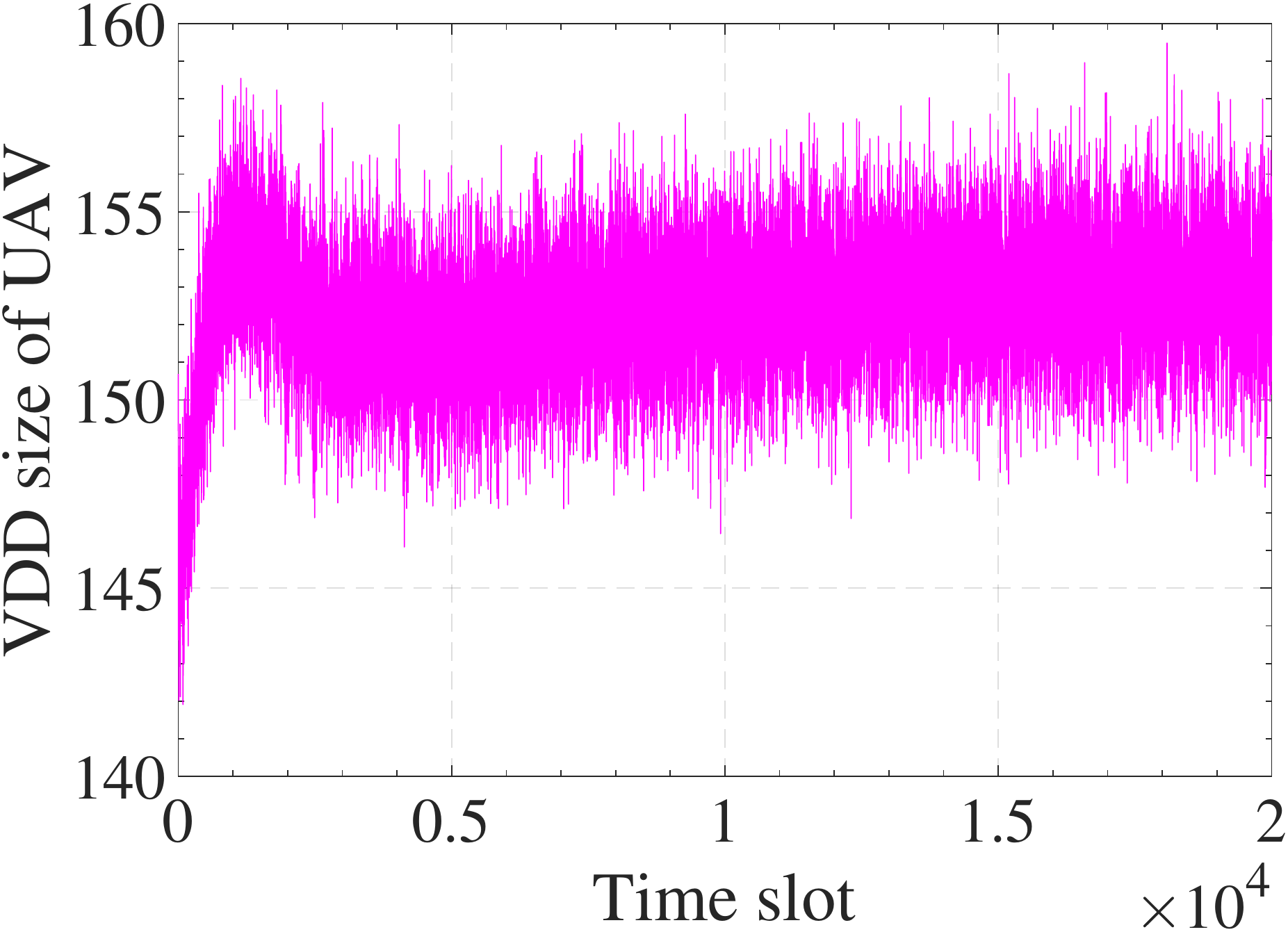}
    \caption{Evolution of UAV's strategy on VDD size using PHC learning under complete information asymmetry.}\label{fig:simu3}
\end{minipage}~~
\begin{minipage}[t]{0.243\textwidth}
\centering
    \includegraphics[height=3.37cm,width=\linewidth]{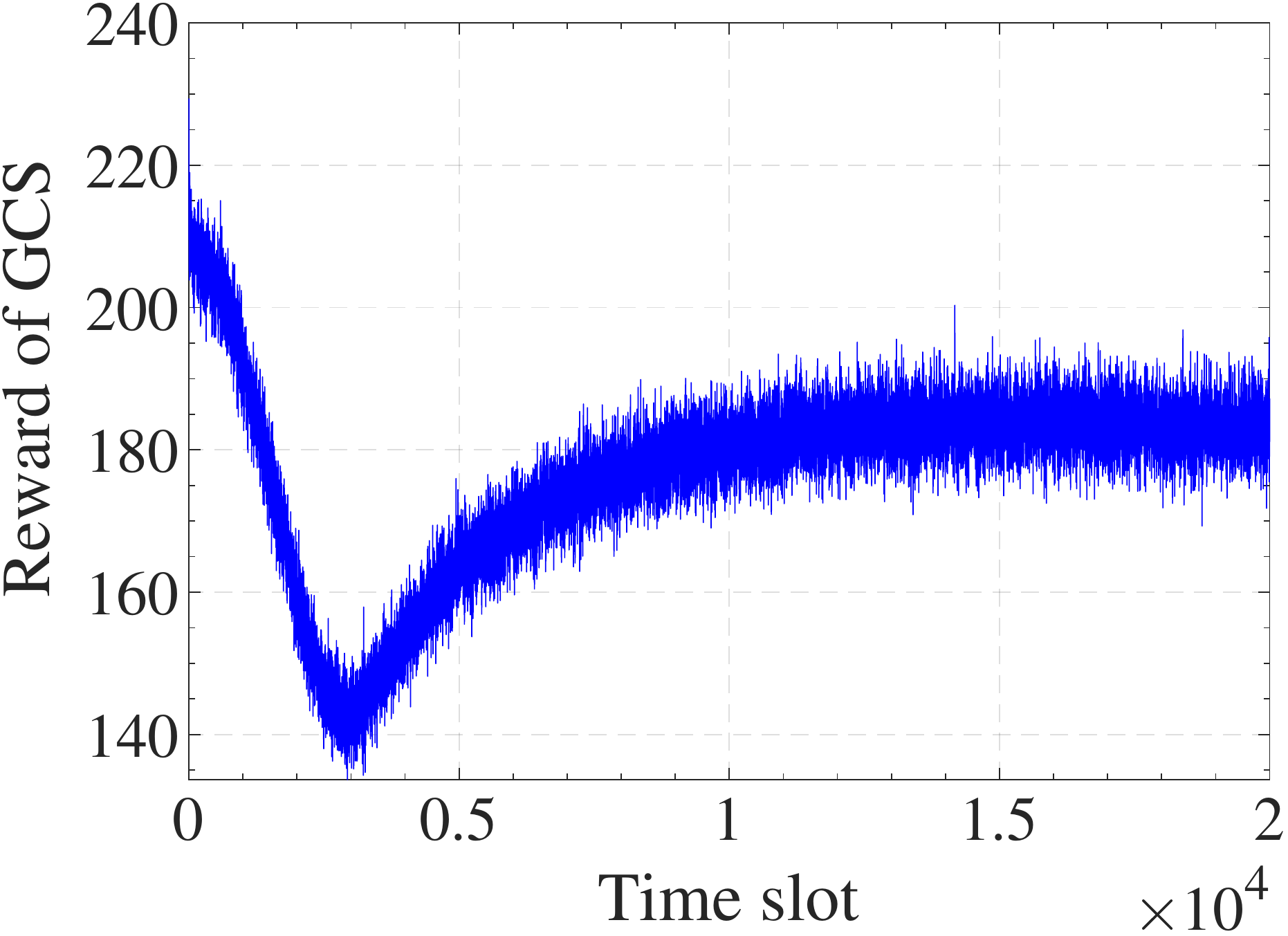}
    \caption{Evolution of GCS's reward strategy using PHC learning under complete information asymmetry.}\label{fig:simu4}
\end{minipage}\vspace{-0.35cm}
\end{figure*}

\vspace{-0.05cm}
We first evaluate the utility of UAV and the defensive effectiveness in Figs.~\ref{fig:simu1}--\ref{fig:simu2}. Here, the \emph{defensive effectiveness} is defined as $\zeta = \frac{\sum_{j\in \mathcal{J}'}{S_j}}{D_G}$, where ${D_G}$ is the VDD requirement of the GCS. Here, we set ${D_G}=800$ bytes.
As seen in Fig.~\ref{fig:simu1}, the UAV's utility remains zero under no information asymmetry. The reason is that GCS intends to maximize its utility while enforcing IR, which is consistent with (\ref{eq:CIoptS})--(\ref{eq:CIoptR}). Moreover, in Fig.~\ref{fig:simu1}, the higher type (i.e., lower marginal VDD cost) brings higher utility to the UAV, which conforms to the monotonicity of the optimal contract. Overall, our proposal attains higher utility for high-type UAVs (with lower marginal VDD cost) than the linear contract, and higher utility for low-type UAVs than the uniform contract.

As shown in Fig.~\ref{fig:simu2}, our proposed scheme under partial information asymmetry outperforms both linear and uniform contracts in terms of higher defensive effectiveness, and its gap with the ideal complete information contract shrinks as the number of UAVs decreases. The reason is that the reward in the uniform and linear contracts is either fixed or linear with the VDD. While the relationship between the optimal reward and optimal VDD size in optimal contracts is nonlinear in our proposal, creating a stronger incentive for UAVs to contribute more VDD and improve defensive effectiveness.

Next, in Figs.~\ref{fig:simu3}--\ref{fig:simu4}, we show the convergence of PHC-based optimal contract for a randomly selected UAV under complete information asymmetry. The evolutions of UAV's strategy on VDD size and GCS's reward strategy via PHC learning are shown in Fig.~\ref{fig:simu3} and Fig.~\ref{fig:simu4}, respectively. As seen in these two figures, both the VDD size and reward in dynamic contracts can converge to stable and optimal values, validating the feasibility of the proposed two-tier PHC learning-based incentive mechanism. In Fig.~\ref{fig:simu3}, the VDD size first increases then converges to a stable state, while the corresponding reward in Fig.~\ref{fig:simu4} first decreases then grows to reach the stable state. The reasons are as follows.
Motivated by the initial high reward of the GCS, the UAV intends to share more VDD to improve its utility. Then, after observing UAV's high VDD contribution, the GCS gradually decreases its reward to increase its utility. After that, the UAV and GCS continuously pursue their maximized utilities by seeking the optimal VDD size and the optimal reward based on their observed system states.

\section{Conclusion}\label{sec:CONSLUSION}
Collaborative defense among UAVs is essential to combat large-scale and advanced attacks on UAV networks. 
By exchanging trapped attack data in local honeypots, this paper has proposed an optimal and feasible incentive mechanism to promote collaborative defense for UAVs. 
A novel honeypot game has been formulated between the GCS and UAVs with distinct types (i.e., VDD cost and communication delay), the solution of which is to design optimal VDD-reward contracts under both partial and complete information asymmetry scenarios.
For partial information asymmetry, we have analytically derived the optimal contracts with contractual feasibility by summarizing UAV's multi-dimensional private information into a one-dimensional metric.
Besides, a two-tier PHC learning-based algorithm has been devised to address the dynamic contract design problem under complete information asymmetry.
Numerical results have validated the effectiveness of the proposed scheme in terms of UAV utility and defensive performance, compared with conventional schemes.
For future work, we plan to study UAV's privacy protection and the trust issues in VDD sharing.

\bibliographystyle{IEEETran}
\bibliography{ref-UAV_honeypot}
\end{document}